\documentstyle[prd,preprint,tighten,aps,eqsecnum,%
amssymb,newlfont,epsfig]{revtex}
\newcommand{\AD}{D}
\begin{document}
\draft
\preprint{
\begin{tabular}{r}
UWThPh-1998-19\\
DFTT 15/98\\
hep-ph/9804421\\
\today
\end{tabular}
}
\title{Four-Neutrino Mixing and Big-Bang Nucleosynthesis}
\author{S.M. Bilenky}
\address{Joint Institute for Nuclear Research, Dubna, Russia, and\\
INFN, Sezione di Torino, and Dipartimento di Fisica Teorica,
Universit\`a di Torino,\\
Via P. Giuria 1, I--10125 Torino, Italy}
\author{C. Giunti}
\address{INFN, Sezione di Torino, and Dipartimento di Fisica Teorica,
Universit\`a di Torino,\\
Via P. Giuria 1, I--10125 Torino, Italy}
\author{W. Grimus and T. Schwetz}
\address{Institute for Theoretical Physics, University of Vienna,\\
Boltzmanngasse 5, A--1090 Vienna, Austria}
\maketitle
\begin{abstract}
We investigate the Big-Bang Nucleosynthesis
constraints on neutrino mixing
in the framework of the two
four-neutrino schemes
that are favored by the results
of neutrino oscillation experiments.
We discuss the implications of these constraints for
terrestrial
short and long-baseline neutrino oscillation experiments
and we present some possibilities of testing them
in these experiments. 
In particular,
we show that from the
Big-Bang Nucleosynthesis constraints it follows that the
$
\stackrel{\scriptscriptstyle (-)}{\nu}_{\hskip-3pt \mu} 
\to
\stackrel{\scriptscriptstyle (-)}{\nu}_{\hskip-3pt \tau}
$
transition is severely suppressed in short-baseline
experiments,
whereas its oscillation amplitude in long-baseline experiments
is of order 1.
We also propose a new parameterization
of the four-neutrino mixing matrix $U$ which
is appropriate
for the schemes under consideration.
\end{abstract}


\section{Introduction}
\label{sec1}

The problem of the masses and mixing of neutrinos
(see Refs. \cite{BP78,BP87,Mohapatra-Pal,CWKim})
is the first priority problem of
neutrino physics.
Many experiments searching for neutrino
oscillations and neutrinoless double-beta decay and
investigating
the high-energy part of the tritium beta-spectrum are going on
or are under preparation.
At present, indications that neutrinos are
massive and mixed
have been found in
solar neutrino experiments
(Homestake \cite{Homestake},
Kamiokande \cite{Kam-sun},
GALLEX \cite{GALLEX},
SAGE \cite{SAGE}
and
Super-Kamiokande \cite{SK-sun,SK}),
in atmospheric neutrino experiments
(Kamiokande \cite{Kam-atm},
IMB \cite{IMB},
Soudan \cite{Soudan}
and
Super-Kamiokande \cite{SK-atm,SK})
and
in the LSND experiment \cite{LSND}.
From the analyses of the data of these experiments
in terms of neutrino oscillations
it follows that there are three different scales
of neutrino mass-squared differences:
\begin{eqnarray}
&
\Delta{m}^2_{\mathrm{sun}}
\sim
10^{-5} \, \mathrm{eV}^2
\, (\mbox{MSW})
\quad \mbox{or} \quad
\Delta{m}^2_{\mathrm{sun}}
\sim
10^{-10} \, \mathrm{eV}^2
\, (\mbox{vac. osc.})
\quad
\mbox{\cite{HL97,FLM97}}
\,,
&
\label{DMsun}
\\
&
\Delta{m}^2_{\mathrm{atm}}
\sim
5 \times 10^{-3} \, \mathrm{eV}^2
\quad
\mbox{\cite{Valencia}}
\,,
&
\label{DMatm}
\\
&
\Delta{m}^2_{\mathrm{SBL}} \sim 1 \, \mathrm{eV}^2
\quad
\mbox{\cite{LSND}}
\,,
&
\label{DMlsnd}
\end{eqnarray}
where
$\Delta{m}^2_{\mathrm{SBL}}$
is the neutrino mass-squared difference
relevant for short-baseline (SBL) experiments,
whose allowed range is determined by the
positive results of the LSND experiment.
The two possibilities for
$\Delta{m}^2_{\mathrm{sun}}$
correspond,
respectively,
to the
MSW \cite{MSW}
and
to the
vacuum oscillation
solutions of the solar neutrino problem.

Thus,
at least four light neutrinos with definite masses
(with only three active flavor
neutrinos, $\nu_e$, $\nu_\mu$, $\nu_\tau$)
must exist in nature
in order to accommodate the results of all neutrino oscillation
experiments.
This means that
there exists at least one non-interacting sterile 
neutrino
\cite{four,BGKP,BGG96,OY96,BGG97a,BGG97b,Barger-Gibbons}.
In this case,
a left-handed neutrino field
$\nu_{{\alpha}L}$
($\alpha=e,\mu,\tau,s$)
is a mixture of the left-handed
components
$\nu_{kL}$
of the four fields of neutrinos
with definite masses
$m_k$ ($k=1,\ldots,4$):
\begin{equation}
\nu_{{\alpha}L}
=
\sum_{k=1}^{4}
U_{{\alpha}k}
\,
\nu_{kL}
\qquad
(\alpha=e,\mu,\tau,s)
\,,
\label{05}
\end{equation}
where $U$ is the $4\times4$ unitary mixing matrix.

In Ref. \cite{BGG96}
we considered the schemes with four massive neutrinos
and we have
shown that from six possible mass spectra of four massive
neutrinos only the following two mass spectra with two pairs
of close masses separated by a gap of about 1 eV
(the ``LSND gap'')
are compatible with all existing data
(see also Ref. \cite{OY96}):
\begin{equation}\label{spectrum}
\mbox{(A)}
\qquad
\underbrace{
\overbrace{m_1 < m_2}^{\mathrm{atm}}
\ll
\overbrace{m_3 < m_4}^{\mathrm{sun}}
}_{\mathrm{SBL}}
\qquad \mbox{and} \qquad
\mbox{(B)}
\qquad
\underbrace{
\overbrace{m_1 < m_2}^{\mathrm{sun}}
\ll
\overbrace{m_3 < m_4}^{\mathrm{atm}}
}_{\mathrm{SBL}}
\,.
\label{AB}
\end{equation}
In scheme A,  
$ \Delta{m}^{2}_{\mathrm{atm}} \equiv m_2^2-m_1^2 $
and
$ \Delta{m}^{2}_{\mathrm{sun}} \equiv m_4^2-m_3^2 $
are the mass-squared differences relevant for atmospheric and solar
neutrino oscillations, respectively,
whereas in scheme B we have
$ \Delta{m}^{2}_{\mathrm{sun}} \equiv m_2^2-m_1^2 $
and
$ \Delta{m}^{2}_{\mathrm{atm}} \equiv m_4^2-m_3^2 $.
In both schemes
$ \Delta{m}^{2}_{\mathrm{SBL}} \equiv m_4^2-m_1^2 $
is relevant
for neutrino oscillations in short-baseline
experiments.

In this paper we study
the compatibility of these two schemes
with an upper bound\footnote{For $N_\nu \geq 4$
the schemes A and B are trivially
compatible with BBN.}
$N_\nu \lesssim 3.9$
(see Ref. \cite{copi})
on the effective number $N_\nu$
of light neutrinos relevant in
Big-Bang Nucleosynthesis (BBN)
(see, for example, Ref. \cite{KT90}).
We derive the constraints on
some elements of the neutrino mixing matrix $U$
that follow from $N_\nu \lesssim 3.9$
(see also Ref. \cite{OY96})
and
we discuss some possibilities to
check these constraints in
future short-baseline (SBL) and long-baseline (LBL)
neutrino oscillation experiments.

Let us notice that
there was recently some discussion in the literature about the validity
of the upper bound
$N_\nu \lesssim 3.9$
(see Ref. \cite{copi}).
This discussion was
initiated by conflicting results of different measurements of the 
deuterium abundance in
high-redshift hydrogen clouds \cite{highD,lowD}.
The situation is not
completely clarified,
but there seems to exist now a strong case  
in favor of $N_\nu \lesssim 3.9$ \cite{copi}.

Let us define the quantities
\begin{equation}\label{c}
c_{\alpha}
\equiv
\sum_{k=1,2} |U_{{\alpha}k}|^2
\qquad
(\alpha=e,\mu,\tau,s)
\,.
\end{equation}
Okada and Yasuda \cite{OY96}
have shown that the bound
$N_\nu \lesssim 3.9$
implies stringent limits on
$c_s$ in scheme A
and
$1-c_s$ in scheme B.
This means that
the solar neutrino deficit is explained 
by small angle MSW $\nu_e\to\nu_s$ 
transitions\footnote{The vacuum oscillation and the
large angle MSW solutions of the solar neutrino problem
are disfavored by the data
in the case of
$\nu_e\to\nu_s$ 
transitions
(see Refs. \cite{KP,petcov,HL97}).}
and the atmospheric neutrino anomaly by
$\nu_\mu\to\nu_\tau$ transitions\footnote{This explanation
of the atmospheric neutrino anomaly
is consistent with the recent results
of the first
long-baseline reactor experiment CHOOZ \cite{CHOOZ}
in which
no indications in favor of
$\bar\nu_e$
disappearance were found.
}
(see Ref. \cite{Valencia}).
In this paper we have reanalyzed the constraints on the
parameter $c_s$
that follow from BBN.
We have obtained more stringent upper bounds on
$c_s$ in scheme A
and
$1-c_s$ in scheme B
than those presented in Ref. \cite{OY96}
and we have investigated
the effects of these
constraints for SBL and LBL neutrino oscillations,
some of which can 
be tested in the near future. 

The plan of the paper is as follows.
In Section \ref{sec2} we
summarize
the formalism of SBL neutrino oscillations in the schemes A and B
(for more details see Refs. \cite{BGKP,BGG96}).
In Section \ref{sec3} we
investigate in detail the dependence of $N_\nu$
on $c_s$.
From the BBN upper bound on $N_\nu$
we obtain upper bounds on
$c_s$ in scheme A
and
$1-c_s$ in scheme B.
In Sections \ref{sec4} and \ref{sec5}
we obtain various general relations between
SBL and LBL oscillation amplitudes and $c_s$.
These relations
can be useful to check
the BBN constraints and to test
the hypothesis of existence of a sterile neutrino.
In Section \ref{sec5} 
we also introduce a new
parameterization of the mixing matrix that is suitable for the schemes
under consideration.

\section{Neutrino oscillations in SBL experiments}
\label{sec2}

In the schemes under consideration
the probability of
$\nu_\alpha\to\nu_\beta$
transitions ($\beta\neq\alpha$)
and the survival probability of $\nu_\alpha$
in SBL experiments
are given by
\cite{BGKP}
\begin{eqnarray}
&&
P_{\nu_\alpha\to\nu_\beta}
=
\frac{1}{2} \, A_{\alpha;\beta} \,
\left( 1 - \cos \frac{ \Delta{m}^2_{\mathrm{SBL}} L }{ 2 p } \right)
\,,
\label{Ptran}
\\
&&
P_{\nu_\alpha\to\nu_\alpha}
=
1
-
\frac{1}{2} \, B_{\alpha;\alpha} \,
\left( 1 - \cos \frac{ \Delta{m}^2_{\mathrm{SBL}} L }{ 2 p } \right)
\,,
\label{Psurv}
\end{eqnarray}
with
the oscillation amplitudes
\begin{eqnarray}
&&
A_{\alpha;\beta}
=
4
\left|
\sum_i
U_{{\beta}i}
\,
U_{{\alpha}i}^*
\right|^2
\,,
\label{Aab}
\\
&&
B_{\alpha;\alpha}
=
4
\left( \sum_i |U_{{\alpha}i}|^2 \right)
\left( 1 - \sum_i |U_{{\alpha}i}|^2 \right)
\,,
\label{Baa}
\end{eqnarray}
where the index $i$ runs over the values $1,2$
or $3,4$.

In the following,
we will consider values of $\Delta{m}^2_{\mathrm{SBL}}$
in the wide range
\begin{equation}
10^{-1} \, \mathrm{eV}^2
\leq
\Delta{m}^2_{\mathrm{SBL}}
\leq
10^{3} \, \mathrm{eV}^2
\,.
\label{widerange}
\end{equation}
The direct comparison of the regions in the
$A_{\mu;e}$--$\Delta{m}^2_{\mathrm{SBL}}$
plane
allowed at 90\% CL by the results of the LSND experiments
with
the regions excluded at 90\% CL by the results of other SBL experiments
indicate a considerably smaller allowed range
for
$\Delta{m}^2_{\mathrm{SBL}}$:
\begin{equation}
0.27 \, \mathrm{eV}^2
\lesssim
\Delta{m}^2_{\mathrm{SBL}}
\lesssim
2.2 \, \mathrm{eV}^2
\,.
\label{LSNDrange}
\end{equation}
The results of the combined analysis
of the data of all SBL experiments
presented in Ref. \cite{FLS97} show that
the lower bound
$ \Delta{m}^2_{\mathrm{SBL}} \gtrsim 0.27 \, \mathrm{eV}^2 $
is robust
(if the LSND indication is correct),
whereas
values of
$\Delta{m}^2_{\mathrm{SBL}}$
larger than
$ 2.2 \, \mathrm{eV}^2 $
maybe are not excluded.

Taking into account the results of solar
and atmospheric neutrino experiments,
from the exclusion curves of SBL disappearance experiments
at each possible value of 
$\Delta{m}^2_{\mathrm{SBL}}$
we have \cite{BGG96}
\begin{eqnarray}
&&
c_e \leq a_e^0
\qquad \mbox{and} \qquad
1 - c_\mu \leq a_\mu^0
\qquad \mbox{in scheme A}
\,,
\label{041}
\\
&&
1 - c_e \leq a_e^0
\qquad \mbox{and} \qquad
c_\mu \leq a_\mu^0
\qquad \mbox{in scheme B}
\,,
\label{042}
\end{eqnarray}
where 
\begin{equation}
a_\alpha^0
\equiv
\frac{1}{2}
\left( 1 - \sqrt{ 1 - B_{\alpha;\alpha}^0 } \, \right)
\qquad
(\alpha=e,\mu)
\label{a0}
\end{equation}
and
$B_{\alpha;\alpha}^0$
is the upper bound for the amplitude
$B_{\alpha;\alpha}$
that can be obtained from the exclusion plots of
SBL reactor and accelerator disappearance experiments.
From the 90\% CL exclusion curves of the
Bugey \cite{Bugey95}
$\bar\nu_e\to\bar\nu_e$
reactor experiment
and of the
CDHS \cite{CDHS84} and CCFR \cite{CCFR84}
$\nu_\mu\to\nu_\mu$
accelerator experiments
it follows that
$a^{0}_{e}$
is small
($ a^{0}_e \lesssim 4 \times 10^{-2} $)
for
$\Delta m_{\mathrm{SBL}}^{2}$
in the range (\ref{widerange})
and
$ a^{0}_\mu \lesssim 10^{-1} $
for
$\Delta m_{\mathrm{SBL}}^{2} \gtrsim 0.5 \, \mathrm{eV}^2$
(see Fig. 1 of Ref. \cite{BBGK96}).

\section{BBN constraints on 4-neutrino mixing}
\label{sec3}

In this section we derive
the BBN constraints on the elements of the mixing matrix
in the two schemes A and B.
We follow the standard arguments 
\cite{kainulainen,barbieri,enqvist91,enqvist92,OY96}
leading to the BBN constraints
on the elements of the mixing matrix $U$
if a sterile neutrino exists.
According to the standard BBN scenario
(see, for example, Ref. \cite{KT90})
a lepton asymmetry can be neglected
and the active neutrinos are in chemical
equilibrium.
Thus, the following considerations are valid for a
temperature range $T_{\mathrm{dec}} \lesssim T \lesssim 100$ MeV
with the decoupling temperature
$T_{\mathrm{dec}} \sim 2$ MeV for electron neutrinos and
$T_{\mathrm{dec}} \sim 4$ MeV for muon and tau neutrinos.
In this temperature region,
the effective potentials of neutrinos
due to coherent forward scattering
in the primordial plasma
are given by \cite{raffelt,enqvist91,enqvist92}
\begin{equation}
V_e \equiv V = -6.02\, G_F\, p \frac{T^4}{M_W^2} \,, \;
V_{\mu,\tau} = \xi V \quad \mbox{and} \quad V_s=0
\end{equation}
where $p$ is the neutrino momentum, which henceforth will be replaced
by its temperature average $\langle p \rangle \simeq 3.15 \:T$, and 
$\xi = \cos^2 \theta_W / (2+\cos^2 \theta_W) \simeq 0.28$.
The propagation of neutrinos\footnote{In the absence of a 
lepton asymmetry the neutrinos and antineutrinos
evolve identically and hence only neutrinos are considered here.} 
is governed by the effective hamiltonian \cite{MSW}
\begin{eqnarray}
H_\nu
\null & \null = \null & \null
\frac{1}{2p}
\, U \, \mathrm{diag}(m_1^2,m_2^2,m_3^2,m_4^2) \, U^{-1}
+ \mathrm{diag}(V,\xi V,\xi V,0)
\nonumber
\\
\null & \null = \null & \null
U' \, \mbox{diag}(E_1,E_2,E_3,E_4) \, U'^{-1}
\,,
\label{Hnu}
\end{eqnarray}
where
$E_1,E_2,E_3,E_4$
are the effective energy eigenvalues of neutrinos in matter
and
$U'$ is the corresponding effective mixing matrix.
The amount of sterile neutrinos present at nucleosynthesis can be
calculated using the differential equation
\cite{kainulainen} 
\begin{equation}\label{master}
\frac{dn_{\nu_s}}{dt}=
\frac{1}{2} \sum_{\alpha=e,\mu,\tau}
\langle P_{\nu_\alpha\to\nu_s} \rangle_{\mathrm{coll}}
\Gamma_{\nu_\alpha} (1-n_{\nu_s}) 
\end{equation}
where $n_{\nu_s}$ is the number density of the sterile neutrino relative
to the number density of an active neutrino in equilibrium and
$\Gamma_{\nu_\alpha}$ are the collision rates of the active neutrinos,
including elastic and inelastic scattering \cite{enqvist92}, given by
\begin{equation}\label{Gamma}
\Gamma_{\nu_e} = 4.0\, G_F^2 T^5 \quad \mbox{and} \quad
\Gamma_{{\nu_\mu},{\nu_\tau}} =B \, \Gamma_{\nu_e}
\quad \mbox{with} \quad B=0.722 \,.
\end{equation}
Let us notice some important characteristics of Eq.(\ref{master}):
\def\theenumi{\roman{enumi}}
\begin{enumerate}
\item
The probabilities for $\nu_\alpha\to\nu_s$ transitions 
are averaged over the collision time 
$t_{\mathrm{coll}} = 1/\Gamma_{\nu_e}$ and
also $n_{\nu_s}$ has to be considered as such an averaged quantity.
\item
This equation should
approximately hold for non-resonant and adiabatic resonant 
neutrino transitions, as long as
the active neutrinos are in chemical equilibrium and 
$t_{\mathrm{osc}} \ll t_{\mathrm{coll}} \ll t_{\mathrm{exp}}$ is
fulfilled.
The characteristic expansion time of the universe
$t_{\mathrm{exp}}$ is given by
$t_{\mathrm{exp}} = 1/H$
where $H$ is the Hubble parameter,
which is related to the temperature $T$ by 
$H=-\dot{T}/T \simeq 0.69\, (T/1\,\mathrm{MeV})^2 \,\mathrm{s}^{-1}$.
\item
The 
condition $t_{\mathrm{osc}} \ll t_{\mathrm{coll}}$ means that neutrino
oscillations have to be fast relative to the
collision time.
The relation
$\Gamma_{\nu_e}/H \simeq 1.2\, (T/1\, \mathrm{MeV})^3$ shows that for
temperatures larger than 2 MeV the collision time is always much
smaller than the expansion time \cite{kainulainen}.
\item
Chemical equilibrium for the active neutrinos 
is maintained at temperatures larger than the
neutrino decoupling temperature even in the presence of sterile neutrinos,
provided the relevant mixing angles in
the primordial plasma are not too large \cite{enqvist92}.
\end{enumerate}

Using the relation
$H=-\dot{T}/T$,
the evolution equation (\ref{master}) for $n_{\nu_s}$
can be rewritten as
\begin{equation}\label{master1}
\frac{dn_{\nu_s}}{dT} = -\frac{1}{2HT}
\sum_{\alpha=e,\mu,\tau}
\langle P_{\nu_\alpha\to\nu_s} \rangle_{\mathrm{coll}}
\Gamma_{\nu_\alpha} (1-n_{\nu_s}) \,.
\end{equation}
Since by definition $N_\nu$
is the effective number of massless neutrino species
at $T_{\mathrm{dec}}$,
in order to get a constraint on the mixing of sterile neutrinos
we need to calculate
the value of $n_{\nu_s}$ at $T_{\mathrm{dec}}$
produced by neutrino oscillations.
With the initial condition
$n_{\nu_s}(T_i)=0$ ($T_i\sim 100$ MeV),
the integration of
Eq.(\ref{master1}) gives \cite{cline}
\begin{equation}\label{int}
n_{\nu_s}(T_{\mathrm{dec}}) = 
1-\exp\left(-\int_{T_{\mathrm{dec}}}^{T_i} \frac{1}{2HT} 
\sum_{\alpha=e,\mu,\tau}
\langle P_{\nu_\alpha\to\nu_s} \rangle_{\mathrm{coll}}
\Gamma_{\nu_\alpha}
dT\right) \equiv 1-\exp(-F) \,.
\end{equation}
Imposing the upper bound
\begin{equation}
n_{\nu_s}(T_{\mathrm{dec}}) \leq \delta N \equiv N_\nu - 3 \,,
\end{equation}
one obtains the condition 
\begin{equation}\label{cond}
F \leq |\ln(1-\delta N)| \,.
\end{equation}

Since $c_e$ is small in scheme A
and
$1-c_e$ is small in scheme B
(see Eqs.(\ref{041}) and (\ref{042})),
they do not have any effect on neutrino oscillations during BBN.
Hence in this Section we use the approximation
$c_e=0$ in scheme A
and the approximation
$1-c_e=0$ in scheme B \cite{OY96}.
In this case,
the elements
$U_{ei}$ and $U_{si}$ $(i=1, \ldots, 4)$
of the mixing matrix can be chosen real and
can be parameterized by
\begin{eqnarray}
(U_{ei}) = (0, 0, c_\theta , s_\theta),\quad
(U_{si}) = (s_\varphi s_\chi , -s_\varphi c_\chi ,
                -c_\varphi s_\theta , c_\varphi c_\theta) 
\quad \mbox{in scheme A}, \\
(U_{ei}) = (c_\theta , s_\theta ,0,0),\quad
  (U_{si}) = (s_\varphi s_\theta , -s_\varphi c_\theta ,
                -c_\varphi s_\chi , c_\varphi c_\chi)
\quad \mbox{in scheme B},
\end{eqnarray}
with the abbreviations
$c_\rho \equiv \cos \rho$ and
$s_\rho \equiv \sin \rho$
for
$\rho=\theta,\chi,\varphi$
and
$0 \leq \varphi \leq \pi/2$.
In both schemes A and B we have
$ c_s = \sin^2 \varphi $.

Then, it can be shown that
\cite{OY96}
\begin{eqnarray}
\sum_{\alpha = e, \mu, \tau}
\langle P_{\nu_\alpha\to\nu_s} \rangle_{\mathrm{coll}}
\Gamma_{\nu_\alpha}
=
\frac{B\, \Gamma_{\nu_e}}{2}
& &
\left\{
\frac{\AD^2 \sin^2 2\chi}
{1 - 2 \AD \xi x_{\mathrm{atm}} \cos 2\chi + \AD^2 (\xi x_{\mathrm{atm}})^2} 
\right.
\nonumber
\\
& &
{} + \left.
\frac{4 c_s (1-c_s)}
{1 - 2 \xi x_{\mathrm{SBL}} (1-2c_s)
+ (\xi x_{\mathrm{SBL}})^2} \right\} \label{P}
\end{eqnarray}
where
\begin{equation}
x_\sigma \equiv \frac{2pV}{\Delta{m}^2_\sigma} =
- 2.18 \times 10^{-7}
\left( \frac{\Delta{m}^2_\sigma}{1\,\mathrm{eV}^2} \right)^{-1}
\left( \frac{T}{1\, \mathrm{MeV}} \right)^6 \qquad
(\sigma = \mathrm{atm, SBL})
\end{equation}
and $p$ is replaced by its temperature average.
With the definitions
\begin{equation}\label{defA}
\AD \equiv c_s \quad \mbox{in scheme A} 
\quad \mbox{and} \quad
\AD \equiv 1 - c_s \quad \mbox{in scheme B}
\,,
\end{equation}
Eq. (\ref{P}) holds in both schemes.
The two terms in Eq. (\ref{P}) correspond to
oscillations due to $\Delta{m}^2_{\mathrm{atm}}$
and $\Delta{m}^2_{\mathrm{SBL}}$, respectively. Oscillations due to
$\Delta{m}^2_{\mathrm{sun}}$ are neglected  in Eq.(\ref{P})
because their
contribution to $F$ (see Eq.(\ref{int})) is at least two orders of
magnitude smaller for $T > T_{\mathrm{dec}}$.

Using the expression (\ref{P}) one can find an analytic form of the
integral (\ref{int}) and the condition (\ref{cond}) gives
the bound
\begin{equation}\label{bound}
920 \left( \frac{\Delta{m}^2_{\mathrm{SBL}}}{1\, \mathrm{eV}^2} \right)^{1/2}
c_s \, \sqrt{1-c_s}
+
33 \left( \frac{\Delta{m}^2_{\mathrm{atm}}}{10^{-2} \, 
\mathrm{eV}^2} \right)^{1/2}
\frac{\sin^2 2\chi}{\sqrt{1+\cos 2\chi}} \, \AD^{3/2}
\leq |\ln(1-\delta N)| \,.
\end{equation}
With the definitions (\ref{defA}) this bound holds for both schemes,
provided the conditions laid down in i.--iv. are fulfilled.

Let us now discuss the implications of the bound (\ref{bound}).
In order to understand if this bound is compatible with a large
$\AD$,
we first consider the possibility that $\AD$ is close to one.
In this case we have
$(U_{\mu 1},U_{\mu 2}) \sim (\cos \chi, \sin \chi)$
in scheme A
and
$(U_{\mu 3},U_{\mu 4}) \sim (\cos \chi, \sin \chi)$
in scheme B.
This means that,
in order to accommodate
the atmospheric neutrino anomaly,
$\sin^2 2\chi$ cannot be small.
However, this is in contradiction
with the inequality (\ref{bound})
because of the second term in the left-hand side.
Therefore, we conclude that the bound (\ref{bound})
implies that $\AD$ is small,
i.e.
$c_s$ is small in scheme A
and
$1-c_s$ is small in scheme B. 

Let us now check the validity of this reasoning
considering the oscillations due to $\Delta{m}^2_{\mathrm{atm}}$
which generate the second term in Eq.(\ref{bound})
and
taking into account
that the method considered here only holds for
non-resonant neutrino oscillations or adiabatic transitions through a
resonance.
Looking at Eq.(\ref{P})
we see that there is a resonant behaviour
of the oscillations due to $\Delta{m}^2_{\mathrm{atm}}$
at the temperature where
the relation
$\AD \xi x_{\mathrm{atm}}=\cos 2\chi$
is satisfied.
Since $x_{\mathrm{atm}}$ is negative, this is only possible
if $\cos 2\chi$ is negative.
However, since we do not know the sign of $\cos 2\chi$,
we have to check if the resonance is crossed
adiabatically in the case of a negative sign.
A measure of the adiabaticity of the resonance is given by 
the parameter
\begin{equation}
q_{\mathrm{atm}} = 
\left| \frac{1}{\Delta E_{\mathrm{atm}}}\, \frac{d\chi'}{dt} 
\right|_{\mathrm{max}}
= 1.23 \times 10^{-4}
\left( \frac{\Delta{m}^2_{\mathrm{atm}}}{10^{-2} \, \mathrm{eV}^2}
\right)^{-1/2}
\frac{\sin 2\chi}{\sqrt{\AD}\, (1+\cos 2\chi)^{3/2}}
\end{equation}
where $\Delta E_{\mathrm{atm}}$ and $\chi'$ are the oscillation
frequency and the effective mixing angle in matter.
If $q_{\mathrm{atm}} \ll 1$
the adiabatic crossing of the
resonance is guaranteed. Clearly, 
for $\AD \sim 1$ and $\sin^2 2\chi \sim 1$ this condition is fulfilled.
Furthermore, oscillations due to $\Delta{m}^2_{\mathrm{atm}}$
occur around the temperature
$T_{\mathrm{atm}} \sim 7 \AD^{-1/6}$ MeV,
where
$\AD |\xi x_{\mathrm{atm}}| \sim 1$.
One can show that for
$\AD \sim 1$
the relation
$t_{\mathrm{osc}} \ll t_{\mathrm{coll}}$
holds for
resonant and non-resonant oscillations due to
$\Delta{m}^2_{\mathrm{atm}}$ and thus the conditions for the validity of
Eq.(\ref{master}) and all relations derived from it are
fulfilled as well.
Consequently, the exclusion of the case $\AD \sim 1$ with the
help of the second term in Eq.(\ref{bound}) is correct.
It is
interesting to note that no conclusions can be drawn from the second
term in Eq.(\ref{bound}) if $\AD \lesssim 0.1$ because
then $T_{\mathrm{atm}} \gtrsim 10$ MeV and
in this temperature region one has $t_{\mathrm{osc}} \gg t_{\mathrm{coll}}$
and therefore no oscillations due to $\Delta{m}^2_{\mathrm{atm}}$ are
possible because of strong quantum damping. 

Let us now examine the implications of the
first term in Eq.(\ref{bound}).
The presence of this term
implies that either $c_s$ or $\sqrt{1-c_s}$
has to be very small.
As we have just shown considering the second term
in Eq.(\ref{bound}),
in scheme A $c_s$ is small.
This implies that
$1-2c_s > 0$
and only non-resonant oscillations are possible
(see the last term in Eq.(\ref{P})).
Hence, Eq.(\ref{bound}) is valid in scheme A and
the first term gives the bound
\begin{equation}\label{bound1A}
c_s \leq 1.1 \times 10^{-3}
\left( \frac{\Delta{m}^2_{\mathrm{SBL}}}{1\, \mathrm{eV}^2} \right)^{-1/2}
|\ln(1-\delta N)| \equiv \mathcal{B}^{(\mathrm{A})}
\,. 
\end{equation}

On the other hand,
considering the second term
in Eq.(\ref{bound}),
we have shown that in scheme B the quantity $1-c_s$ is small.
Therefore, in scheme B we have
$1-2c_s < 0$
and a resonance occurs at the temperature
$T_{\mathrm{res}} = 16 (\Delta{m}^2_{\mathrm{SBL}} / 1\,\mathrm{eV}^2)^{1/6}
|1-2c_s|^{1/6}$ MeV,
where $\xi x_{\mathrm{SBL}}=1-2c_s$.
One can show that this resonance is not passed adiabatically, hence
the conditions for the validity of Eq.(\ref{master}) are not fulfilled and
the first term of Eq.(\ref{bound}) does not apply.
In this case the amount of sterile neutrinos produced at
the resonance through non-adiabatic transitions
can be calculated with the formula \cite{enqvistres,shi}
\begin{equation}
n_{\nu_s} =  \frac{1}{2}- \left(\frac{1}{2} - P_{LZ}\right)
\cos 2\varphi_b\cos 2\varphi_a \simeq 1 - P_{LZ}
\,,
\label{MSWres}
\end{equation}
where $\varphi_b$ ($\varphi_a$) is the effective mixing angle before
(after) the resonance.
The last approximation in Eq.(\ref{MSWres})
follows from the fact that
$\cos 2\varphi_b \simeq 1$
because of the high
effective neutrino potential before the
resonance and
$\cos 2\varphi_a \simeq \cos 2\varphi = 1-2c_s \simeq -1$ for
small $1-c_s$.
The quantity
$P_{LZ} = \exp(-Q)$ is the Landau--Zener probability \cite{zener}
and $Q$ is given by \cite{enqvistres,enqvist92,shi}
\begin{eqnarray}
Q
\equiv
\frac{\pi^2}{2} \, \frac{\delta t}{t_{\mathrm{osc}}}
\null & \null = \null & \null
\frac{2\pi}{\hbar}\frac{c_s(1-c_s)}{|1-2c_s|}
\left|V \left( \frac{dx_{\mathrm{SBL}}}{dt}
\right)^{-1} \right|_{T_{\mathrm{res}}}
\nonumber
\\
\null & \null \simeq \null & \null
9.2\times 10^4 \, \frac{c_s(1-c_s)}{|1-2c_s|^{3/2}}
\left(\frac{\Delta{m}^2_{\mathrm{SBL}}} {1\:\mathrm{eV}^2}\right)^{1/2}
\,,
\label{Q}
\end{eqnarray}
where $\delta t$ is the half-width of the resonance defined in 
Ref. \cite{enqvist91}.
Demanding again that $n_{\nu_s} \leq \delta N$ one obtains
the condition $Q \leq |\ln(1-\delta N)|$ which gives the bound
\begin{equation} \label{bound1B}
1 - c_s \leq 1.1 \times 10^{-5}
\left( \frac{\Delta{m}^2_{\mathrm{SBL}}}{1\:\mathrm{eV}^2}\right)^{-1/2}
|\ln(1-\delta N)| \equiv \mathcal{B}^{(\mathrm{B})} 
\end{equation}
in scheme B.
The linear approximation of the potential used to derive the
Landau--Zener formula is valid
if the potential changes slowly in the resonance region.
At the
resonance one obtains
$|V/(dV/dt)|/\delta t \sim |1-2c_s|/2\sqrt{c_s(1-c_s)}$
which is large for small $1-c_s$.
Hence the
linear approximation should apply.

In the following table we show the values of the bounds
$\mathcal{B}^{(\mathrm{A})}$ and $\mathcal{B}^{(\mathrm{B})}$
obtained from
Eqs.(\ref{bound1A}) and (\ref{bound1B})
and the LSND lower bound
$\Delta{m}^2_{\mathrm{SBL}} \gtrsim 0.27 \, \mathrm{eV}^2$
(see Eq.(\ref{LSNDrange}))
for different values of $\delta N$:
\begin{equation}\label{table}
\begin{array}{cc|cccccccc}
&\delta N & 0.9 & 0.8 & 0.7 & 0.6 & 0.5 & 0.4 & 0.3 & 0.2 \\ \hline
\mbox{scheme A:}\quad & \mathcal{B}^{(\mathrm{A})} /10^{-3}\:\:
\rule[3mm]{0mm}{2mm} 
&\: 4.8 & 3.4 & 2.5 & 1.9 & 1.4 & 1.1 & 0.7 & 0.5 \\
\mbox{scheme B:}\quad & \mathcal{B}^{(\mathrm{B})}/10^{-5}\:\:
&\: 4.9 & 3.4 & 2.6 & 1.9 & 1.5 & 1.1 & 0.8 & 0.5
\end{array}\end{equation}
The bounds (\ref{bound1A}) for the non-resonant oscillations in
scheme A and (\ref{bound1B}) for the resonant case in scheme B are in
rough agreement with the corresponding numerical integrations
of the evolution equations of an
ensemble of oscillating neutrinos
which take into account 2-neutrino
oscillations \cite{enqvist92,shi}.

Let us now summarize our findings.
The constraints on the parameter
$c_s$ in scheme A (Eq.(\ref{bound1A}))
and $1-c_s$ in scheme B (Eq.(\ref{bound1B}))
has been derived
in both schemes from the oscillations due to $\Delta{m}^2_{\mathrm{SBL}}$
at temperatures of the order $T \sim 16$ MeV where 
$\xi |x_{\mathrm{SBL}}| \sim 1$.
The second term in Eq.(\ref{bound}),
which is generated by the oscillations due to $\Delta{m}^2_{\mathrm{atm}}$,
only serves to exclude large $\AD$.
The transitions of active into
sterile neutrinos
due to $\Delta{m}^2_{\mathrm{SBL}}$
are non-resonant
in scheme A,
whereas they are resonant in scheme B,
leading to a stronger constraint on $1-c_s$ than
the constraint on $c_s$ in scheme A
(see table (\ref{table})).
We have obtained stronger bounds than Ref. \cite{OY96} 
because we have taken into account the
complete collision rates (\ref{Gamma}) presented in Ref. \cite{enqvist92},
which are nearly an order of magnitude larger than those used in
Ref. \cite{OY96}. Furthermore, 
our method shows that the dependence of
the bound on
$c_s$ in scheme A and $1-c_s$ in scheme B
on the allowed number of neutrino species $N_\nu$
is logarithmic \cite{cline}.

We want to
emphasize that our calculation 
of the amount of sterile neutrinos brought into equilibrium by
oscillations before the onset of BBN
is based upon equation (\ref{master}) in
scheme A and upon the Landau--Zener approximation in scheme B.
Therefore, the numbers
given in table (\ref{table}) should be viewed as
order of magnitude estimates of the upper bounds on the
parameter $c_s$ in scheme A and $1-c_s$ in scheme B.

Concluding this section,
we would like to make some comments
on the non-standard
BBN scenario with a non-zero lepton asymmetry presented in Ref. \cite{foot}.
There it
has been shown that for a certain range of the mixing parameters
and for resonant transitions due to $\Delta{m}^2_{\mathrm{SBL}}$ 
a lepton asymmetry
as small as $10^{-10}$ can be amplified up to $10^{-2}$ and in this way
transitions at a lower temperature (like the transitions 
due to $\Delta{m}^2_{\mathrm{atm}}$ in the schemes under
consideration)
can be prevented. Such a scenario
would not change our bound for $1-c_s$ in scheme B, but could have
some effect on our considerations for scheme A. It $c_s$ is large in
scheme A, a resonance due to $\Delta{m}^2_{\mathrm{SBL}}$ occurs and,
if a large lepton asymmetry is generated, the oscillations due to  
$\Delta{m}^2_{\mathrm{atm}}$ are suppressed and the second term on the
left-hand side of Eq.(\ref{bound}) is not valid. Hence a large
parameter $c_s$ cannot be excluded. Therefore, contrary to our
discussion in this section, in this case we would have two possibilities for
scheme A: either $c_s$ is small and has to obey the bound
(\ref{bound1A}) or $c_s$ is close to one and
$1-c_s$ has to obey the same
bound as given in Eq.(\ref{bound1B}) for scheme B.
However,
judging from the results of Ref. \cite{foot} it has yet to be shown
that with $c_s$ close to one 
a realistic scenario meeting all experimental constraints
can actually be achieved in scheme A and we leave this 
problem to future investigation.

\section{Terrestrial neutrino oscillation experiments
and sterile neutrino mixing}
\label{sec4}

As we have shown in the previous Section,
BBN suggests that the parameter
$c_s$
in scheme A
($1-c_s$ in scheme B)
is small.
In this Section we present some
possibilities to obtain information
on the parameter $c_s$ from future results of SBL and LBL
neutrino oscillation experiments.
We consider explicitly scheme A,
but the same results are valid in scheme B after
the replacement of
$c_s$ with $1-c_s$.

Let us start with the consideration of
$\nu_\mu\to\nu_\tau$
oscillations,
that are presently searched for in
the CHORUS \cite{CHORUS} and NOMAD \cite{NOMAD} experiments and will 
be searched for in the COSMOS \cite{COSMOS} experiment.
From Eq.(\ref{Aab}),
for the amplitude of
$\nu_\mu\to\nu_\tau$
oscillations we have the upper bound
\begin{equation}
A_{\mu;\tau}
\leq
4 (1-c_\mu) (1-c_\tau)
\,.
\label{06}
\end{equation}
The unitarity of the mixing matrix implies that
\begin{equation}
\sum_{\alpha} c_\alpha = 2
\label{07}
\end{equation}
and Eq.(\ref{06}) can be rewritten as
\begin{equation}
A_{\mu;\tau}
\leq
4 (1-c_\mu) [c_e+c_s-(1-c_\mu)]
\,.
\label{080}
\end{equation}
Since the quantity
$1-c_\mu$
satisfies the inequality (\ref{041}),
if
\begin{equation}
c_e+c_s \geq 2 a_\mu^0
\label{09}
\end{equation}
we have
\begin{equation}
A_{\mu;\tau}
\leq
4 a_\mu^0 (c_e+c_s-a_\mu^0)
\leq
4 a_\mu^0 (a_e^0+c_s)
\,.
\label{08}
\end{equation}
On the other hand,
if the inequality
\begin{equation}
c_e+c_s \leq 2 a_\mu^0
\label{10}
\end{equation}
is satisfied,
for the amplitude
$A_{\mu;\tau}$
we have
\begin{equation}
A_{\mu;\tau}
\leq
(c_e+c_s)^2
\leq
4 (a_\mu^0)^2
\,.
\label{11}
\end{equation}

As it is seen from Eqs.(\ref{08}) and (\ref{11}),
if it will be found that the amplitude $A_{\mu;\tau}$
satisfies the inequality
\begin{equation}
A_{\mu;\tau}
>
4 (a_\mu^0)^2
\,,
\label{111}
\end{equation}
it will mean that the parameter $c_s$ must satisfy the inequality
\begin{equation}
c_s \geq 2 a_\mu^0 - a_e^0
\label{112}
\end{equation}
and is in general not small.
In Fig.\ref{fig1} we have plotted the values of
$4 (a_\mu^0)^2$
(solid curve)
and
$B_{\mu;\mu}^0$
(dashed curve)
obtained from the 90\% CL exclusion curves of the
CDHS \cite{CDHS84} and CCFR \cite{CCFR84} experiments
in the range (\ref{widerange}) of $\Delta{m}^2_{\mathrm{SBL}}$,
that covers the LSND-allowed region.
The dashed curve representing $B_{\mu;\mu}^0$
constitutes an upper bound for
$A_{\mu;\tau}$
due to the conservation of probability.
Therefore,
if the inequality
(\ref{111}) is satisfied,
$A_{\mu;\tau}$
must lie between the solid and dashed curves.
In Fig.\ref{fig1} we have also plotted
the most recent exclusion curve
presented by the CHORUS collaboration \cite{CHORUS98}
(dash-dotted curve),
the expected final sensitivity of the CHORUS \cite{CHORUS}
and NOMAD \cite{NOMAD} experiments
(dash-dot-dotted curve)
and the expected sensitivity of the COSMOS \cite{COSMOS} experiment
(dotted curve).
One can see that a substantial part of the region in which
the inequality
(\ref{111}) is satisfied is already ruled out by the
present CHORUS exclusion curve,
a large part will be excluded
when the final sensitivities of the CHORUS and NOMAD experiments
will be reached
and almost all the region will be excluded
if the COSMOS experiment will not find
$\nu_\mu\to\nu_\tau$
oscillations.

In Fig.\ref{fig2} we have plotted the values of
$2 a_\mu^0 - a_e^0$
obtained from the 90\% CL exclusion curves of the
Bugey \cite{Bugey95},
CDHS \cite{CDHS84} and CCFR \cite{CCFR84} experiments.
One can see that for the values of $\Delta{m}^2_{\mathrm{SBL}}$
in which
$ a_e^0 < 2 a_\mu^0 $,
if the inequality
(\ref{111}) is satisfied
$c_s$
must be large and incompatible with the BBN upper bound.
On the other hand,
for the values of $\Delta{m}^2_{\mathrm{SBL}}$
in which
$ a_e^0 > 2 a_\mu^0 $,
$c_s$ does not need
to be large
(the inequality (\ref{09}) is satisfied because
$c_e$ can be larger than $2a_\mu^0$)
and there is no contradiction with the BBN upper bound
on $c_s$
even if the inequality
(\ref{111}) is satisfied.
If $A_{\mu;\tau}$
will be measured and found to
satisfy the inequality (\ref{111}),
also $\Delta{m}^2_{\mathrm{SBL}}$
will be determined
with some accuracy.
In this case, depending on the allowed
range of $\Delta{m}^2_{\mathrm{SBL}}$,
it will be possible to decide if a small $c_s$
compatible with the BBN upper bound is excluded or not.
Note, however, that for the LSND range (\ref{LSNDrange}) of
$\Delta{m}^2_{\mathrm{SBL}}$
the parameter $c_s$ must be large if the inequality (\ref{111}) 
is satisfied.

Furthermore,
if some future experiment will
measure a value of
$A_{\mu;\tau}$
such that
Eq.(\ref{111})
is satisfied,
then the inequality (\ref{08})
will imply in addition the lower bound
\begin{equation}
c_s
\geq
\frac{ A_{\mu;\tau} }{ 4 a_\mu^0 } - a_e^0
\,,
\label{081}
\end{equation}
which is more stringent than the lower bound (\ref{112}) if
$ A_{\mu;\tau} > 8 (a_\mu^0)^2 $.

Let us now consider the possibility that
a future experiment will measure
a small value for the amplitude $A_{\mu;\tau}$,
i.e. such that the inequality (\ref{11}) is satisfied.
This measurement will not give any information on
the value of $c_s$.
In this case,
the BBN bound derived in Section \ref{sec3}
implies that 
$c_s\simeq0$
and for the amplitude
$A_{\mu;\tau}$
we have the upper bound
\begin{equation}
A_{\mu;\tau}
\leq
(a_e^0)^2
\,.
\label{113}
\end{equation}
The numerical values of this upper bound in the range
(\ref{widerange}) of $\Delta{m}^2_{\mathrm{SBL}}$
obtained from the 90\% CL exclusion curve of the Bugey \cite{Bugey95}
experiment
is shown in Fig.\ref{fig3}
(solid curve),
together with
the most recent exclusion curve
presented by the CHORUS collaboration \cite{CHORUS98}
(dash-dotted curve),
the expected final sensitivity of the CHORUS \cite{CHORUS}
and NOMAD \cite{NOMAD} experiments
(dash-dot-dotted curve)
and the expected sensitivity of the COSMOS \cite{COSMOS} experiment
(dotted curve).

Thus,
we have seen that the investigation of
$\nu_\mu\to\nu_\tau$
oscillations in the region of $\Delta{m}^2_{\mathrm{SBL}}$
that includes the LSND-allowed region could allow to obtain
information on the value of the parameter $c_s$
that is important for BBN.

Let us consider now
the possibility to obtain information on the
parameter $c_s$ from the data of inclusive
$\nu_\mu\to\nu_\mu$
and
$\bar\nu_e\to\bar\nu_e$
experiments.
From Eqs.(\ref{Baa}) and (\ref{041})
it follows that 
\begin{eqnarray}
c_e
\null & \null = \null & \null
\frac{1}{2}
\left( 1 - \sqrt{ 1 - B_{e;e} } \right)
\,,
\label{181}
\\
1-c_\mu
\null & \null = \null & \null
\frac{1}{2}
\left( 1 - \sqrt{ 1 - B_{\mu;\mu} } \right)
\,.
\label{182}
\end{eqnarray}
Furthermore,
from Eq.(\ref{07}) we have
\begin{equation}
c_s \geq (1-c_\mu) - c_e
\,.
\label{183}
\end{equation}
From Eqs.(\ref{181})--(\ref{183})
we obtain the inequality
\begin{equation}
c_s
\geq
\frac{1}{2}
\left( \sqrt{ 1 - B_{e;e} } - \sqrt{ 1 - B_{\mu;\mu} } \right)
\,.
\label{184}
\end{equation}
The amplitude
$B_{e;e}$
is small in the whole range (\ref{widerange})
of $\Delta{m}^2_{\mathrm{SBL}}$.
If the parameter $c_s$ is small as suggested by BBN,
also the amplitude
$B_{\mu;\mu}$
must be small.
The existing data do not exclude, however,
large values of $B_{\mu;\mu}$
for
$ \Delta{m}^2_{\mathrm{SBL}} \lesssim 0.3 \, \mathrm{eV}^2 $.
Hence,
our analysis shows that
the investigation of inclusive
$\nu_\mu\to\nu_\mu$
transitions
for
$ \Delta{m}^2_{\mathrm{SBL}} \lesssim 0.3 \, \mathrm{eV}^2 $
would be
interesting for the check of the constraint that follows from BBN.

Let us now consider the amplitude $A_{e;s}$
of
$\nu_e\to\nu_s$
oscillations.
From the upper bound
\begin{equation}
A_{e;s}
\leq
4 \, c_e \, c_s
\leq
4 \, a_e^0 \, c_s
\,,
\label{es1}
\end{equation}
which follows from Eqs.(\ref{c}), (\ref{Aab}) and (\ref{041}),
it is clear that the BBN bound on $c_s$ implies that
$A_{e;s}$
is extremely small.
In this case
the amplitudes measured in SBL $\nu_e$ disappearance experiments
and
SBL $\nu_e\to\nu_\mu$ and $\nu_e\to\nu_\tau$
appearance experiments
are related by
\begin{equation}
B_{e;e} \simeq A_{\mu;e} + A_{e;\tau}
\,.
\label{es4}
\end{equation}
The experimental check of this relation
allows to test the BBN bound on $c_s$.
Indeed,
if SBL $\nu_e$ disappearance experiments
will find neutrino oscillations with a lower limit
$B_{e;e}^{(\mathrm{min})}$
for the amplitude
$B_{e;e}$
such that
$ B_{e;e}^{(\mathrm{min})} > A_{\mu;e}^{(\mathrm{max})} + A_{e;\tau}^{(\mathrm{max})} $,
where
$A_{\mu;e}^{(\mathrm{max})}$ and $A_{e;\tau}^{(\mathrm{max})}$
are the experimental upper limit for
$A_{\mu;e}$ and $A_{e;\tau}$,
respectively,
it will mean that
$A_{e;s}>0$.
In this case,
taking into account that
$ B_{e;e} = A_{\mu;e} + A_{e;\tau} + A_{e;s} $
and that for small
$B_{e;e}$
we have
$c_e \simeq B_{e;e}/4 $,
the inequality
$ A_{e;s} \leq 4 \, c_e \, c_s $
gives the lower bound
\begin{equation}
c_s
\geq
\frac{ A_{e;s} }{ 4 \, c_e }
\simeq
1 - \frac{ A_{\mu;e} + A_{e;\tau} }{ B_{e;e}  }
\geq
1 -
\frac{ A_{\mu;e}^{(\mathrm{max})} + A_{e;\tau}^{(\mathrm{max})} }
{ B_{e;e}^{(\mathrm{min})}  }
\,.
\label{es2}
\end{equation}
If, for example,
it will be found that\footnote{These
values are compatible with the present upper bound
for $B_{e;e}$,
which is given by the results of the Bugey \cite{Bugey95} experiment,
and with the present allowed range of
$A_{\mu;e}$,
which is
given by the results of the LSND \cite{LSND}
and Bugey experiments
($ B_{e;e} \lesssim 4 \times 10^{-2} $
and
$
2 \times 10^{-3} \lesssim
A_{\mu;e}
\lesssim 4 \times 10^{-2}
$
for $\Delta{m}^2_{\mathrm{SBL}}$
in the LSND range (\ref{LSNDrange})).
Notice that the allowed range of
$A_{\mu;e}$
will be checked in the near future by
KARMEN \cite{KARMEN}
and other experiments
\cite{LSNDcheck}.
Furthermore,
information on
$A_{e;\tau}$
could be obtained in the future
in experiments using a
$\nu_e$ beam
from a muon collider
(see Refs. \cite{Geer,Mohapatra97}).}
$B_{e;e}^{(\mathrm{min})}\simeq10^{-2}$,
$A_{\mu;e}^{(\mathrm{max})}\simeq5\times10^{-3}$
and
$A_{e;\tau}^{(\mathrm{max})}{\ll}A_{\mu;e}^{(\mathrm{max})}$,
Eq.(\ref{es2})
will imply the lower bound
$ c_s \gtrsim 0.5 $,
which is incompatible with the BBN upper bound.

CP violation in long-baseline neutrino oscillation experiments can
also be used to check the BBN constraint on $c_s$,
though in practice this possibility is
more remote.
It has been shown in Ref. \cite{BGG97b} that
the absolute value of the CP-violating parameter 
$I_{\mu\tau} = 4\,\mbox{Im}\!\left[
U_{\mu 1} \, U_{\tau 1}^* \, U_{\mu 2}^* \, U_{\tau 2} \right] $
(in scheme A,
whereas
$I_{\mu\tau} = 4\,\mbox{Im}\!\left[
U_{\mu 3} \, U_{\tau 3}^* \, U_{\mu 4}^* \, U_{\tau 4} \right] $
in scheme B)
could be as large as
$2/3\sqrt{3}$,
which is the maximal value allowed by the
unitarity of the mixing matrix.
With the method discussed in
Ref. \cite{BGG97b}
one can show that
\begin{equation}\label{CP1}
|I_{\mu\tau}| \leq 
(c_e+c_s)\sqrt{1-c_e-c_s}
\end{equation}
for small $c_e+c_s$.
From the upper bound
$c_e \leq a_e^0$
(see Eq.(\ref{041}))
and the BBN upper bound
$c_s \leq \mathcal{B}^{(\mathrm{A})}$
(see Eq.(\ref{bound1A}))
we have
\begin{equation}\label{CP2}
|I_{\mu\tau}| \leq 
(a_e^0+\mathcal{B}^{(\mathrm{A})})\sqrt{1-a_e^0-\mathcal{B}^{(\mathrm{A})}}
\,.
\end{equation}
Consequently,
taking into account that
$a_e^0$ is small,
the BBN upper bound on $c_s$
implies that CP violation
in the $\nu_\mu\to\nu_\tau$ channel
is suppressed.
On the other hand,
if CP violation will be observed in LBL
$\nu_\mu\to\nu_\tau$
experiments,
using Eq.(\ref{CP1})
it will be possible to set a lower bound
for $c_s$.
Analogous results are valid in scheme B
with the replacements
$ c_e \to 1-c_e $,
$ c_s \to 1-c_s $
and
$ \mathcal{B}^{(\mathrm{A})} \to \mathcal{B}^{(\mathrm{B})} $.

\section{A parameterization of the 4-neutrino mixing matrix}
\label{sec5}

In this Section we propose a parameterization of the
$4\times4$ mixing matrix $U$ which is suitable for
the two schemes A and B that allow to accommodate all 
existing neutrino oscillation data.
In these schemes the neutrinos $\nu_1$, $\nu_2$
and the neutrinos $\nu_3$, $\nu_4$
give separate contributions to all observables.
Thus,
it is natural to consider separately
the elements $U_{{\alpha}i}$ with $i=1,2$
and
the elements $U_{{\alpha}k}$ with $k=3,4$.
In the following we will stick to scheme A for the actual presentation.
The formulas in scheme B can be obtained with the exchanges 
$1 \leftrightarrows 3$ and $2 \leftrightarrows 4$ of the columns of $U$.

We take into account that all
observable transition probabilities are invariant under the
phase transformation
\begin{equation}
U_{{\alpha}j}
\to
e^{i x_\alpha}
\,
U_{{\alpha}j}
\,
e^{i y_j}
\label{401}
\end{equation}
where $x_\alpha$ and $y_j$ are arbitrary parameters.
Therefore,
$U_{ek}$ with $k=3,4$
and
$U_{{\mu}i}$ with $i=1,2$
can always be considered as real vectors.
Let us write them as
\begin{eqnarray}
&&
U_{ek}
=
r_e \, v_{k-2}
\qquad
(k=3,4)
\,,
\label{402}
\\
&&
U_{{\mu}i}
=
p_\mu \, w_i
\qquad
(i=1,2)
\,,
\label{403}
\end{eqnarray}
with
\begin{equation}
r_e
=
\sqrt{ \sum_{k=3,4} |U_{ek}|^2 }
\quad \mbox{and} \quad
p_\mu
=
\sqrt{ \sum_{i=1,2} |U_{{\mu}i}|^2 }
\,.
\label{404}
\end{equation}
The unit vectors
$v$ and $w$
can be written in the form
\begin{equation}\label{unit}
v = (\cos \theta, \sin \theta)
\quad \mbox{and} \quad
w =(\cos \gamma, \sin \gamma)
\,.
\end{equation}
Let us introduce also the orthogonal unit vectors
\begin{equation}\label{orth}
v^\bot = (-\sin \theta, \cos \theta)
\quad \mbox{and} \quad
w^\bot = (-\sin \gamma, \cos \gamma)
\,.
\end{equation}
Then the vectors
$U_{{\alpha}i}$ with $\alpha\neq{e}$ and $i=3,4$
can be expanded over the orthonormal basis
$\{ v$, $v^\bot \}$
and the vectors
$U_{{\alpha}i}$ with $\alpha\neq\mu$ and $i=1,2$
can be expanded over the orthonormal basis
$\{ w$, $w^\bot \}$.
As a result,
we have
the parameterization
\begin{equation}\label{Uv}
U = \left(
\begin{array}{ll}
p_e w + q_e w^\bot & r_e v \\
p_\mu w            & r_\mu v + s_\mu v^\bot \\
p_\tau w + q_\tau w^\bot & r_\tau v + s_\tau v^\bot \\
p_s w + q_s w^\bot & r_s v + s_s v^\bot
\end{array} \right)
\end{equation}
for the mixing matrix. 
Using the invariance under the transformation (\ref{401}),
we can choose
the parameters $q_\tau$ and $s_s$ as real.

From Eqs.(\ref{c}) and (\ref{404})
it is 
obvious that the parameters $r_e$ and $p_\mu$ are connected,
respectively,
with the parameters
$c_e$ and $c_\mu$
by the relations
\begin{equation}
r_e = \sqrt{1-c_e},
\quad
p_\mu = \sqrt{c_\mu}
\,.
\label{411}
\end{equation}
Furthermore,
from the unitarity of $U$
it follows that the parameters $r_\mu$ and $p_e$ 
are related by 
\begin{equation}
r_\mu = -p_e^* \, \sqrt{\frac{c_\mu}{1-c_e}}
\,.
\label{412}
\end{equation}
Similarly, we can think of $q_\tau$ and $s_s$ being fixed by the unit
length of the lines of $U$ and five of the remaining complex
parameters by the orthogonality of different lines. Counting the
number of remaining real parameters in $U$ we obtain 10 versus 
9 physical real parameters in a 4$\times$4 unitary mixing matrix. A
careful inspection of the residual phase freedom reveals that one of
the complex parameters 
$p_e$, $q_e$, $r_\mu$, $s_\mu$, $p_\tau$, $r_\tau$, $s_\tau$,
$p_s$, $q_s$ or $r_s$ can be chosen real in addition.

The significance of the parameters in the matrix (\ref{Uv})
shows up
by considering
the amplitudes of neutrino 
oscillations in disappearance and appearance SBL experiments,
which are directly connected with the parameters
$p_\alpha$ and $r_\alpha$.
In fact, we have
\begin{equation}\label{BB}
B_{e;e} = 4 r_e^2 (1-r_e^2), \quad
B_{\mu;\mu} = 4 p_\mu^2 (1-p_\mu^2)
\end{equation}
and
\begin{equation}\label{AA}
A_{\mu;e} = 4 |p_e|^2 p_\mu^2 \,, \quad A_{e;\tau} = 4 |r_\tau|^2 r_e^2 \,,
\quad A_{\mu;\tau} = 4 |p_\tau|^2 p_\mu^2 \,.
\end{equation}
The SBL oscillation amplitudes do not contain the angles $\theta$ and
$\gamma$.
This is connected 
with the fact that SBL amplitudes are determined by products of vectors 
in the $1,2$ space or in the $3,4$ space
(see Eqs.(\ref{Aab}) and (\ref{Baa})).
However,
information on the angle $\gamma$ can be obtained from the results of
atmospheric and LBL neutrino experiments,
whereas
the results of solar neutrino experiments give
information on the angle $\theta$.

In the four-neutrino scheme A
the parameters $p_e$, $q_e$ are small because
$c_e = |p_e|^2 + |q_e|^2$
is small
(see Eq.(\ref{041})).
Let us now investigate if it is possible to obtain bounds on the parameters
$|p_s|^2$ and $|q_s|^2$
from measurable quantities.
We will use the unitarity of the mixing matrix and
we will work in the approximation
$c_e=0$.
In this case,
the parameterization (\ref{Uv}) of the mixing matrix
reduces to
\begin{equation}\label{Uv1}
U \simeq \left(
\begin{array}{ll}
0                        & v \\
p_\mu w                  & s_\mu v^\bot \\
p_\tau w + q_\tau w^\bot & s_\tau v^\bot \\
p_s w + q_s w^\bot       & s_s v^\bot
\end{array} \right)
\end{equation}
and
the amplitude of
$\nu_\mu\to\nu_\tau$
oscillations is given by
(see Eq.(\ref{Aab}))
\begin{equation}
A_{\mu;\tau} = 4 |s_\mu|^2 |s_\tau|^2 
\,.
\label{060}
\end{equation}
Furthermore, taking into account that
\begin{eqnarray}
|p_\tau|^2 +  |q_\tau|^2 + |s_\tau|^2 & = & 1
\,,
\label{061}
\\
|q_\tau|^2 +  |q_s|^2 & = & 1
\,,
\label{062}
\end{eqnarray}
we have
\begin{equation}
|q_s|^2 = |s_\tau|^2 + |p_\tau|^2 \geq |s_\tau|^2
\,.
\label{063}
\end{equation}
From Eqs.(\ref{060}) and (\ref{063})
and taking into account that
$ |s_\mu|^2 = 1 - c_\mu \leq a_\mu^0 $
(see Eq.(\ref{041})),
we obtain
\begin{equation}
|q_s|^2 \geq \frac{A_{\mu;\tau}}{4a_\mu^0}
\,.
\label{064}
\end{equation}
This lower bound for
$|q_s|^2=c_s-|p_s|^2$,
obtained in the approximation $a_e^0=0$,
is consistent with the lower bound
(\ref{081}) on $c_s$.
From the comparison of
these two lower bounds
it is clear that if $c_s$ is large,
its value approximately coincides with $|q_s|^2$,
whereas $|p_s|^2$
is small if $a^0_\mu$ is small.
Indeed,
it is possible to obtain an upper bound for the parameter $|p_s|^2$
using the unitarity relation
\begin{equation}
p_\mu \, p_s^* + s_\mu \, s_s^* = 0
\,.
\label{065}
\end{equation}
Taking into account that
\begin{eqnarray}
&&
|s_\mu|^2 = 1 - p_\mu^2
\,,
\label{066}
\\
&&
|s_s|^2 = 1 - |p_s|^2 - |q_s|^2
\,,
\label{067}
\end{eqnarray}
we find from Eq.(\ref{065}) 
\begin{equation}
\frac{ |p_s|^2 }{ 1 - |q_s|^2 }
=
1 - p_\mu^2
\,.
\label{068}
\end{equation}
Since
$ 1 - p_\mu^2 =  1 - c_\mu \leq a_\mu^0 $
(see Eq.(\ref{041})),
we obtain the upper bound
\begin{equation}
|p_s|^2 \leq a_\mu^0
\,.
\label{069}
\end{equation}
Therefore, if the experimental upper bound
for the amplitude of SBL $\nu_\mu\to\nu_\mu$ transitions is small,
$|p_s|^2$ is small as well.

Note that the observation of SBL inclusive
$\nu_\mu\to\nu_\mu$ transitions
together with the observation of SBL
$\nu_\mu\to\nu_\tau$
and
$\nu_\mu\to\nu_e$
(LSND \cite{LSND})
transitions
would allow to get information
on the transitions of muon neutrinos into sterile states.
In fact,
because of the unitarity of the mixing matrix,
for the amplitude $A_{\mu;s}$ of
$\nu_\mu\to\nu_s$
transitions we have
\begin{equation}
A_{\mu;s}
=
B_{\mu;\mu}
-
A_{\mu;e}
-
A_{\mu;\tau}
\,.
\label{070}
\end{equation}
Hence,
we think that
it is important to continue the search for neutrino
oscillations in the inclusive
$\nu_\mu\to\nu_\mu$
channel,
especially in the region of small $\Delta{m}^2_\mathrm{SBL}$,
below $ 0.3 \, \mathrm{eV}^2 $,
which has not been explored so far.
Let us consider the possibility that
the right side of Eq.(\ref{070})
will be measured to be different from zero,
i.e., that
$\nu_\mu\to\nu_s$
transitions occur in SBL experiments.
In the scheme under consideration the amplitude of
these transitions is given by
\begin{equation}
A_{\mu;s}
=
4 \, p_\mu^2 \, |p_s|^2
\,.
\label{071}
\end{equation}
Taking into account that
$ p_\mu^2 = c_\mu $,
Eqs.(\ref{182}), (\ref{070}) and (\ref{071}) imply that
\begin{equation}
|p_s|^2
=
\frac{ B_{\mu;\mu} - A_{\mu;e} - A_{\mu;\tau} }
{ 2 \left( 1 + \sqrt{ 1 - B_{\mu;\mu} } \right) }
\,.
\label{072}
\end{equation}

Finally, if it will be found that not only
$\nu_e\to\nu_e$
SBL transitions but also
$\nu_\mu\to\nu_\mu$
SBL transitions are strongly suppressed,
it will mean that
the only oscillation channel involving sterile neutrinos
which may be not suppressed is
the channel $\nu_\tau\to\nu_s$.
Therefore,
in this case it will be possible to obtain
information on transitions of active neutrinos into sterile states
only if
$\nu_\tau$ beams
(see Ref. \cite{nutau_beams})
will be available.
In this case $|p_s|^2\simeq0$ and
\begin{equation}
B_{\tau;\tau}
=
A_{\tau;s}
=
4 \, |q_s|^2 \left( 1 - |q_s|^2 \right)
\,.
\label{073}
\end{equation}

Let us now consider the possibility that
not only
$p_e$, $q_e$
but also $p_s$, $q_s$
are small,
as follows from the smallness of
$c_s = |p_s|^2 + |q_s|^2$
implied by BBN if
$N_\nu \lesssim 3.9$,
as shown in Section \ref{sec3}.
In this case it is possible to expand all the quantities
$p_\alpha$, $q_\alpha$, $r_\alpha$ and $s_\alpha$
in powers of the small parameters
$p_e$, $q_e$, $p_s$, $q_s$,
up to order
$c_e$, $c_s$, $\sqrt{c_e c_s}$.
This means that in this approximation it is
possible\footnote{Actually,
this is always possible because of unitarity,
but in general the relations are very complicated.}
to express all elements of $U$ in terms of the parameters
$p_e$, $q_e$, $p_s$, $q_s$, $\theta$ and $\gamma$.
Note that one of the phases of
the small complex parameters is unphysical and can be transformed away
reaching thus the number of 9 physically independent real parameters in $U$.
Performing this program we obtain
\begin{equation}\label{elements}
\begin{array}{c}
|p_\mu|^2 \simeq 1 - |p_e|^2 - |p_s|^2,
\qquad
r_\mu \simeq -p_e^*,
\qquad 
s_\mu \simeq -p_s^*,
\\
p_\tau \simeq -p_e q_e^* - p_s q_s^*,
\qquad
|q_\tau|^2 \simeq 1 - |q_e|^2 - |q_s|^2,
\qquad
r_\tau \simeq -q_e^*,
\qquad
s_\tau \simeq -q^*_s,
\\
r_s \simeq -p_e^* p_s - q_e^* q_s,
\qquad 
|s_s|^2 \simeq 1-c_s.
\end{array}
\end{equation}
In this approximation, Eqs.(\ref{BB}) and (\ref{AA}) become
\begin{equation}\label{BB1}
B_{e;e} \simeq 4 (|p_e|^2 + |q_e|^2), \quad
B_{\mu;\mu} \simeq 4 (|p_e|^2 + |p_s|^2)
\end{equation}
and
\begin{equation}\label{AA1}
A_{\mu;e} \simeq 4\, |p_e|^2, \quad A_{e;\tau} \simeq 4 |q_e|^2, \quad 
A_{\mu;\tau} \simeq 4 |p_e q_e^* + p_s q_s^*|^2
\,. 
\end{equation}
Therefore we have
\begin{equation}\label{BAA}
B_{e;e} \simeq A_{\mu;e} + A_{e;\tau} \,.
\end{equation}
The absolute value of $p_s$ is given by
\begin{equation}\label{ps}
|p_s|^2 \simeq \frac{1}{4} ( B_{\mu;\mu} - A_{\mu;e} ) \,.
\end{equation}
In the absence of $\nu_\tau$ disappearance experiments
(see Eq.(\ref{073})), the determination
of the parameter $|q_s|$ needs
further information from LBL experiments.
This is not an artifact of the expansion
in Eq.(\ref{elements}).
The reason is that $B_{\mu;\mu}$ and
$A_{\mu;\tau}$ are small if $|p_s|$ is small and this is true
even if $|q_s|$ is large,
i.e. $c_s$ is large.
This explains the limitations in our search
for tests of the BBN constraint on $c_s$
derived in Section \ref{sec3}.
On the other hand,
in the present approximation, for the determination of
the values of
$|p_e|$, $|q_e|$ and $|p_s|$ in principle  
the results of SBL experiments alone are sufficient.

The
free parameter $|p_e|^2$ is fixed by
a measurement of the SBL $\nu_\mu \leftrightarrows \nu_e$
oscillation amplitude.
With the result of the LSND experiment and the negative result
of all other SBL neutrino oscillation experiments
$|p_e|^2$ is approximately bounded by 
$0.5 \times 10^{-3} \lesssim |p_e|^2 \lesssim 10^{-2}$.
If the BBN constraint on $c_s$ is valid,
then $\nu_\mu\to\nu_\tau$
transitions are significantly suppressed in SBL neutrino oscillation
experiments with an oscillation amplitude of order $10^{-3}$ or smaller.
Furthermore, Eq.(\ref{BAA}) shows that
the oscillation amplitude $A_{e;\tau}$ is at most close to 0.1.

In the discussion in Section \ref{sec3}
we have seen that the BBN bound $N_\nu \lesssim 3.9$ implies that
$c_s \ll 10^{-2}$.
In this case we have
\begin{equation}\label{relation}
B_{\mu;\mu} \simeq A_{\mu;e} \quad \mbox{and} \quad
A_{\mu;\tau} \simeq \frac{1}{4} \, A_{\mu;e} \, A_{e;\tau} \,.
\end{equation}

Inspecting Eq.(\ref{elements}) we note that in scheme A the quantity 
$1-c_\mu$ is now of the same
order of magnitude as $c_e$.
As mentioned in the previous Section, 
this is particularly interesting for
$\Delta{m}^2_{\mathrm{SBL}} \lesssim 0.3$ eV$^2$
where the bound on $B_{\mu;\mu}$
from the $\nu_\mu$ disappearance experiments
is not very stringent and, consequently,
in this $\Delta{m}^2_{\mathrm{SBL}}$ region $B_{\mu;\mu}$
should be much smaller than the present
upper experimental upper bound.
In addition, also $1-c_\tau$ is of
the same order of magnitude,
showing that
at zeroth-order in the expansion over
the small quantities
$c_e$ and $c_s$,
the mixing matrix is given by
\begin{equation}
U \simeq \left(
\begin{array}{ll}
0      & v \\
w      & 0 \\
w^\bot & 0 \\
0      & v^\bot
\end{array} \right)
\label{mix1}
\end{equation}
Therefore, we have
\begin{eqnarray}
&&
\nu_e
\simeq
\cos\theta \, \nu_3 + \sin\theta \, \nu_4
\,,
\qquad
\nu_s
\simeq
- \sin\theta \, \nu_3 + \cos\theta \, \nu_4
\,,
\nonumber
\\
&&
\nu_\mu
\simeq
\cos\gamma \, \nu_1 + \sin\gamma \, \nu_2
\,,
\qquad
\nu_\tau
\simeq
- \sin\gamma \, \nu_1 + \cos\gamma \, \nu_2
\,.
\label{mix2}
\end{eqnarray}
Hence,
the $\nu_e,\nu_s$ -- $\nu_3,\nu_4$
and
$\nu_\mu,\nu_\tau$ -- $\nu_1,\nu_2$
sectors are decoupled
and the oscillations of solar
and atmospheric neutrinos are independent.
Furthermore,
since only the small
mixing angle MSW solutions of the solar neutrino problem
seems to be allowed by the data
in the case of
$\nu_e\to\nu_s$ 
transitions
(see Refs. \cite{KP,petcov,HL97}),
at zeroth-order in the expansion over
the small quantities
$c_e$, $c_s$ and $\theta$
we have
\begin{equation}
\nu_e \simeq \nu_3
\,,
\qquad
\nu_s \simeq \nu_4
\,.
\label{mix3}
\end{equation}
Similar conclusions
are valid in scheme B
with $c_\alpha$ replaced by $1-c_\alpha$
and
$1,2 \leftrightarrows 3,4$.

Let us now discuss LBL neutrino oscillations.
Keeping only the dominant terms we get 
\begin{equation}
P^{(\mathrm{LBL})}_{
\stackrel{\scriptscriptstyle(-)}{\nu}_{\hskip-3pt \mu} \to 
\stackrel{\scriptscriptstyle(-)}{\nu}_{\hskip-3pt \tau}      }
\simeq
1 - P^{(\mathrm{LBL})}_{
\stackrel{\scriptscriptstyle(-)}{\nu}_{\hskip-3pt \mu} \to 
\stackrel{\scriptscriptstyle(-)}{\nu}_{\hskip-3pt \mu}      }
\simeq \sin^2 (2\gamma) \, \sin^2 (\phi/2)
\end{equation}
where $\phi \simeq \Delta{m}^2_{21} L/2p$ in scheme A and
$\Delta{m}^2_{43} L/2p$ in scheme B.
This shows
that there must be significant LBL $\nu_\mu\to\nu_\tau$
oscillations because of the sizeable $\nu_\mu$ disappearance
known from the atmospheric neutrino anomaly,
which implies that $\sin^2 (2\gamma)$ is large.
For
$\nu_\mu\to\nu_e$ and $\nu_e\to\nu_\tau$ transitions 
matter effects have to be taken
into account.
The upper bounds on these transition probabilities
derived in Ref. \cite{BGG97a}
are of order $10^{-2}$.
If $c_s$ is small,
as follows from the BBN constraint,
the ``matter-stable'' bounds on
$
P^{(\mathrm{LBL})}_{
\stackrel{\scriptscriptstyle(-)}{\nu}_{\hskip-3pt \mu} \to 
\stackrel{\scriptscriptstyle(-)}{\nu}_{\hskip-3pt e} }
$
and
$
P^{(\mathrm{LBL})}_{
\stackrel{\scriptscriptstyle(-)}{\nu}_{\hskip-3pt e} \to 
\stackrel{\scriptscriptstyle(-)}{\nu}_{\hskip-3pt \tau} }
$
derived in Ref. \cite{BGG97a}
are improved for small values of
$\Delta{m}^2_{\mathrm{SBL}}$
because of the inequalities
\begin{equation}
c_{\mu} \;, c_{\tau} \geq 1-a^0_e-c_s
\,,
\end{equation}
leading to
\begin{equation}
P^{(\mathrm{LBL})}_{
\stackrel{\scriptscriptstyle(-)}{\nu}_{\hskip-3pt \mu} \to 
\stackrel{\scriptscriptstyle(-)}{\nu}_{\hskip-3pt e} }
\;,
P^{(\mathrm{LBL})}_{
\stackrel{\scriptscriptstyle(-)}{\nu}_{\hskip-3pt e} \to 
\stackrel{\scriptscriptstyle(-)}{\nu}_{\hskip-3pt \tau} }
\leq 2 a^0_e(1-a^0_e) + c_s(1-2a^0_e)
\,.
\label{LBL1}
\end{equation}
Hence, the BBN constraint on $c_s$ imply that
\begin{equation}
P^{(\mathrm{LBL})}_{
\stackrel{\scriptscriptstyle(-)}{\nu}_{\hskip-3pt \mu} \to 
\stackrel{\scriptscriptstyle(-)}{\nu}_{\hskip-3pt e} }
\;,
P^{(\mathrm{LBL})}_{
\stackrel{\scriptscriptstyle(-)}{\nu}_{\hskip-3pt e} \to 
\stackrel{\scriptscriptstyle(-)}{\nu}_{\hskip-3pt \tau} }
\lesssim 2 a^0_e(1-a^0_e)
\,.
\label{LBL2}
\end{equation}
Obviously,
the observation of a violation of these inequalities
in LBL neutrino oscillation experiments
would imply that the BBN constraint on $c_s$
is not valid.

\section{Conclusions}
\label{sec6}

In this paper we have focused our discussion
on the two four-neutrino schemes A and B
(see Eq.(\ref{AB}))
which are compatible with
the results of all
neutrino oscillation experiments
\cite{BGG96,OY96}.
As shown in Ref. \cite{OY96}, 
the bound $N_\nu \lesssim 3.9$
for the effective number $N_\nu$ of light neutrinos 
before the onset of BBN
implies a stringent upper bound on 
$c_s \equiv \sum_{k=1,2} |U_{sk}|^2$
in scheme A and on $1-c_s$ in scheme B.
Our limits are more stringent than those of Ref. \cite{OY96}
because we
used the complete collision rates presented in Ref. \cite{enqvist92}.
Thus,
in scheme A
we have obtained 
$c_s \lesssim 5 \times 10^{-3}$.
We have demonstrated that in scheme B the limit derives from
resonant transitions due to $\Delta{m}^2_{\mathrm{SBL}}$,
leading to an upper bound for $1-c_s$
that is more stringent than the one for $c_s$ in scheme A:
$1-c_s \lesssim 5 \times 10^{-5}$.

The validity
of these limits from BBN
depends to a certain extent
on the results of the measurements
of the primordial deuterium abundance which are
controversial at the moment \cite{copi}.
We have therefore emphasized the
importance of terrestrial experiments to get information on the value
of $c_s$ in order to check the BBN constraints.
In Section \ref{sec4}
we have worked out
lower bounds on $c_s$ in scheme A ($1-c_s$ in scheme B) involving
the following observables in SBL experiments:
\begin{enumerate}
\item
the $\nu_\mu \to \nu_\tau$ oscillation amplitude
$A_{\mu;\tau}$;
\item
the oscillation amplitude $B_{\mu;\mu}$
in $\nu_\mu$ disappearance experiments
for
$\Delta{m}^2_{\mathrm{SBL}} \lesssim 0.3$ eV$^2$;
\item
the oscillation amplitudes
$B_{e;e}$
in reactor disappearance experiments
and 
$A_{\mu;e}$ and $A_{e;\tau}$
in accelerator experiments.
\end{enumerate}

To proceed further,
in Section \ref{sec5} we have introduced a parameterization of the 4
$\times$ 4 neutrino mixing matrix $U$ (\ref{Uv}) which is particularly 
suited for the schemes A and B. In this
parameterization the quantity $c_s$ is given by 
$c_s = |p_s|^2+|q_s|^2$ in scheme A ($1-c_s = |p_s|^2+|q_s|^2$ in
scheme B),
where $p_s$ and $q_s$ are in general complex parameters in
$U$. The parameterization (\ref{Uv}) has lead us to the following
observations:
\begin{enumerate}
\item[(a)] The parameter $|p_s|$ can be determined in SBL
$\nu_\mu\to\nu_{e,\mu,\tau}$ oscillation experiments.
\item[(b)] For the determination of $|q_s|$ in SBL oscillation
experiments it is necessary to employ a $\nu_\tau$ neutrino beam and
perform a disappearance experiment.
\item[(c)] For $N_\nu \lesssim 3.9$ the SBL oscillation amplitude
$A_{\mu;\tau}$ is at most of order $10^{-3}$ whereas the LBL
oscillation amplitude must be of order one in the same channel.
\end{enumerate}

As already pointed out in Ref. \cite{OY96},
the BBN constraint $N_\nu \lesssim 3.9$
leads to a 4-neutrino mixing picture in which,
apart from small corrections of the order of
the small quantities $c_e$ and $c_s$,
the mixing matrix $U$
is composed by the two decoupled sectors
$\nu_e,\nu_s$ -- $\nu_3,\nu_4$
and
$\nu_\mu,\nu_\tau$ -- $\nu_1,\nu_2$
in scheme A
($\nu_e,\nu_s$ -- $\nu_1,\nu_2$
and
$\nu_\mu,\nu_\tau$ -- $\nu_3,\nu_4$
in scheme B).
Consequently,
the oscillations of solar and atmospheric neutrinos are independent.
The solar neutrino problem is solved with 
$\nu_e\to\nu_s$
oscillations governed by the mixing angle $\theta$
and
the atmospheric neutrino anomaly
is explained with
$\nu_\mu\to\nu_\tau$
oscillations governed by the mixing angle $\gamma$ (see Eq.(\ref{Uv})
for the definition of these angles).
Concerning the solar neutrino deficit, it was shown that
a vacuum oscillation solution involving sterile neutrinos is
rather disfavored by the data (see Refs. \cite{KP,petcov}).
Therefore,
the specific prediction of a small mixing angle MSW solution
of the solar neutrino problem with only
$\nu_e\to\nu_s$ transitions 
serves as an indirect check
of the schemes A and B under consideration
and of the BBN bound.
This prediction
implies a characteristic strong distortion of 
the $^8$B $\nu_e$ spectrum and a
day-night asymmetry
in the neutral current event rate at SNO \cite{krastev}.
Experimentally,
this prediction can be checked 
independently from solar model calculations
(see Ref. \cite{BG}).

If indeed the BBN constraints on the 4-neutrino mixing matrix $U$ are
confirmed then the biggest gap in our knowledge of $U$, the sterile
neutrino mixing, is considerably narrowed. In this paper we have
shown that it is possible to test the ensuing mixing matrix to some
extent in terrestrial experiments. The most striking feature in this
context is the suppression of all SBL oscillation amplitudes.

\begin{figure}[h]
\refstepcounter{figure}
\label{fig1}
FIG.\ref{fig1}.
Plot in the
$A_{\mu;\tau}$--$\Delta{m}^2_{\mathrm{SBL}}$
plane
of the curve
$4 (a_\mu^0)^2$
(solid line).
For
$A_{\mu;\tau}$
to the right of this curve (see Eq.(\ref{111}))
the lower bound (\ref{112})
for $c_s$ applies.
The dashed curve representing $B_{\mu;\mu}^0$
constitutes an upper bound for
$A_{\mu;\tau}$
due to the conservation of probability.
The values of
$a_\mu^0$
and
$B_{\mu;\mu}^0$
have been
obtained from the 90\% CL exclusion curves of the
CDHS \cite{CDHS84} and CCFR \cite{CCFR84} experiments.
The
dash-dotted curve, dash-dot-dotted and dotted curves
represent, respectively,
the most recent exclusion curve
presented by the CHORUS collaboration \cite{CHORUS98},
the expected final sensitivity of the CHORUS \cite{CHORUS}
and NOMAD \cite{NOMAD} experiments
and the expected sensitivity of the COSMOS \cite{COSMOS} experiment.
\end{figure}

\begin{figure}[h]
\refstepcounter{figure}
\label{fig2}
FIG.\ref{fig2}.
Plot in the $c_s$--$\Delta{m}^2_{\mathrm{SBL}}$
plane
of the lower limit
$2 a_\mu^0 - a_e^0$
for $c_s$
that follows from the inequality
(\ref{111}).
The values of
$a_e^0$ and $a_\mu^0$
have been obtained from the 90\% CL exclusion curves of the
Bugey \cite{Bugey95},
CDHS \cite{CDHS84} and CCFR \cite{CCFR84} experiments.
\end{figure}

\begin{figure}[h]
\refstepcounter{figure}
\label{fig3}
FIG.\ref{fig3}.
Plot in the
$A_{\mu;\tau}$--$\Delta{m}^2_{\mathrm{SBL}}$
plane
of the upper bound
$(a_e^0)^2$
(solid curve)
for
$A_{\mu;\tau}$
(see Eq.(\ref{113}))
in the case of a very small $c_s$.
The values of $a_e^0$ have been
obtained from the 90\% CL exclusion curve of the Bugey \cite{Bugey95}
experiment.
The
dash-dotted curve, dash-dot-dotted and dotted curves
are the same as in Fig.\ref{fig1}.
\end{figure}

%\end{document}

\newpage

\begin{minipage}[p]{0.95\textwidth}
\begin{center}
\mbox{\epsfig{file=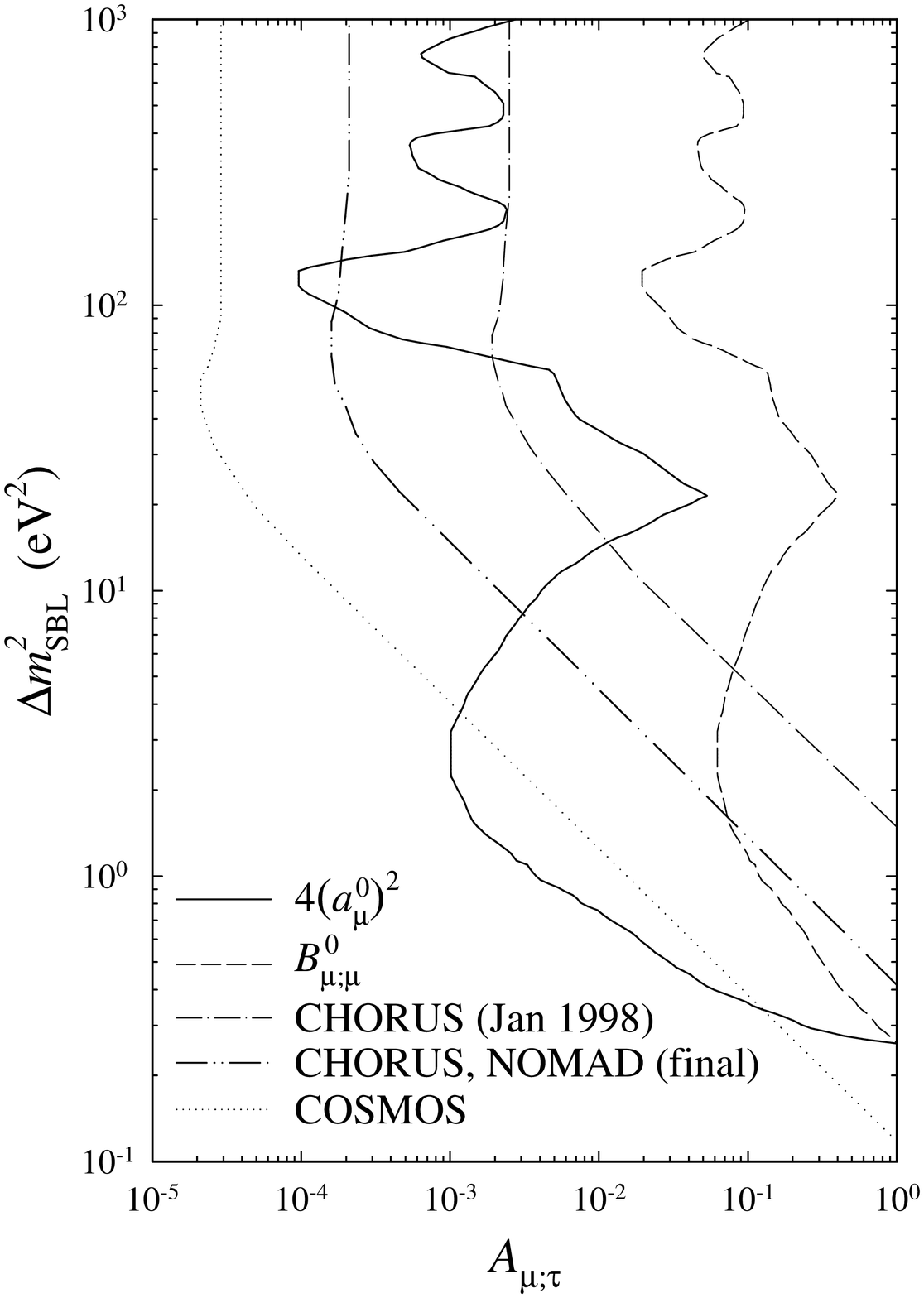,height=0.95\textheight}}
\end{center}
\end{minipage}
\begin{center}
\Large Figure \ref{fig1}
\end{center}

\newpage

\begin{minipage}[p]{0.95\textwidth}
\begin{center}
\mbox{\epsfig{file=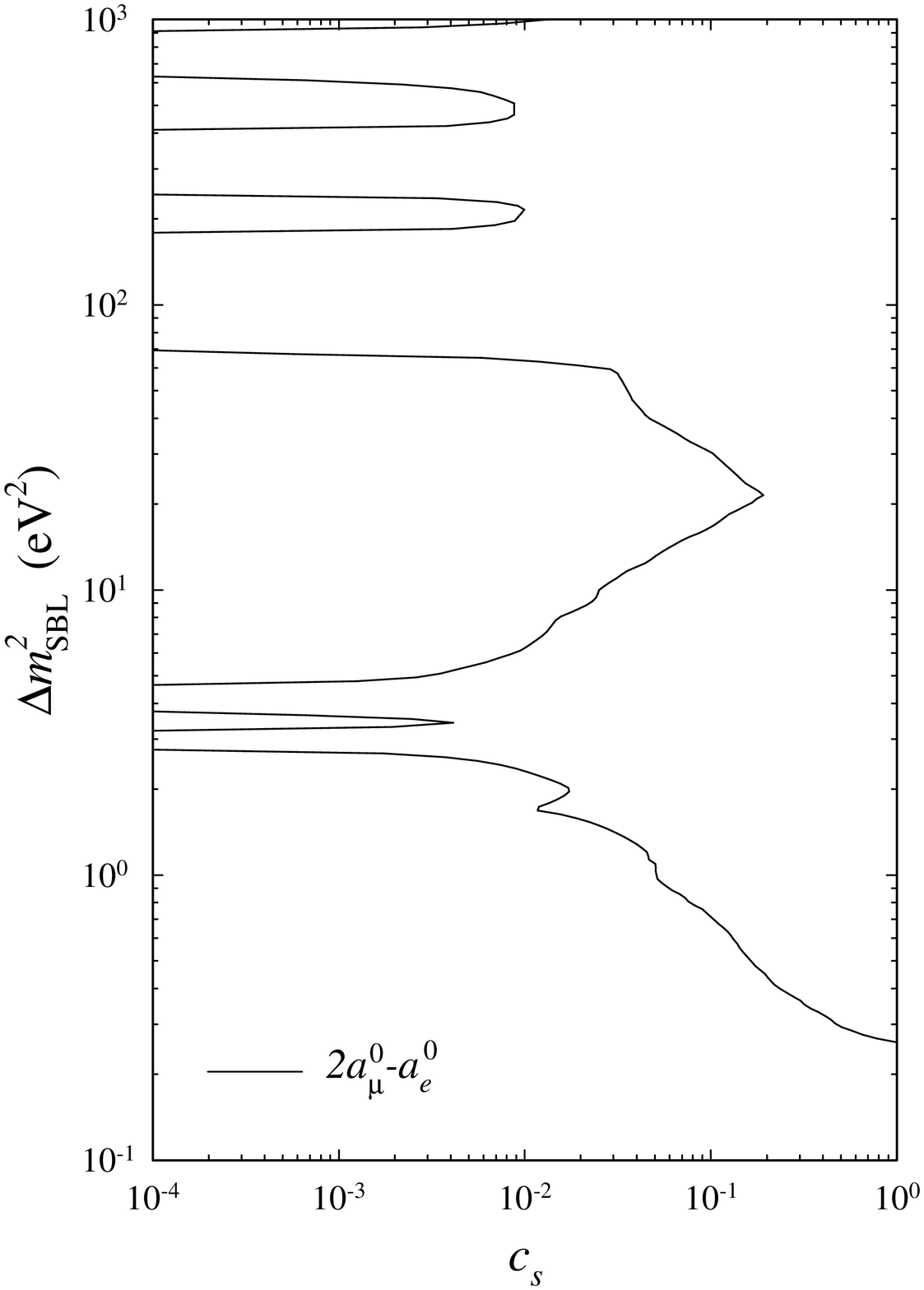,height=0.95\textheight}}
\end{center}
\end{minipage}
\begin{center}
\Large Figure \ref{fig2}
\end{center}

\newpage

\begin{minipage}[p]{0.95\textwidth}
\begin{center}
\mbox{\epsfig{file=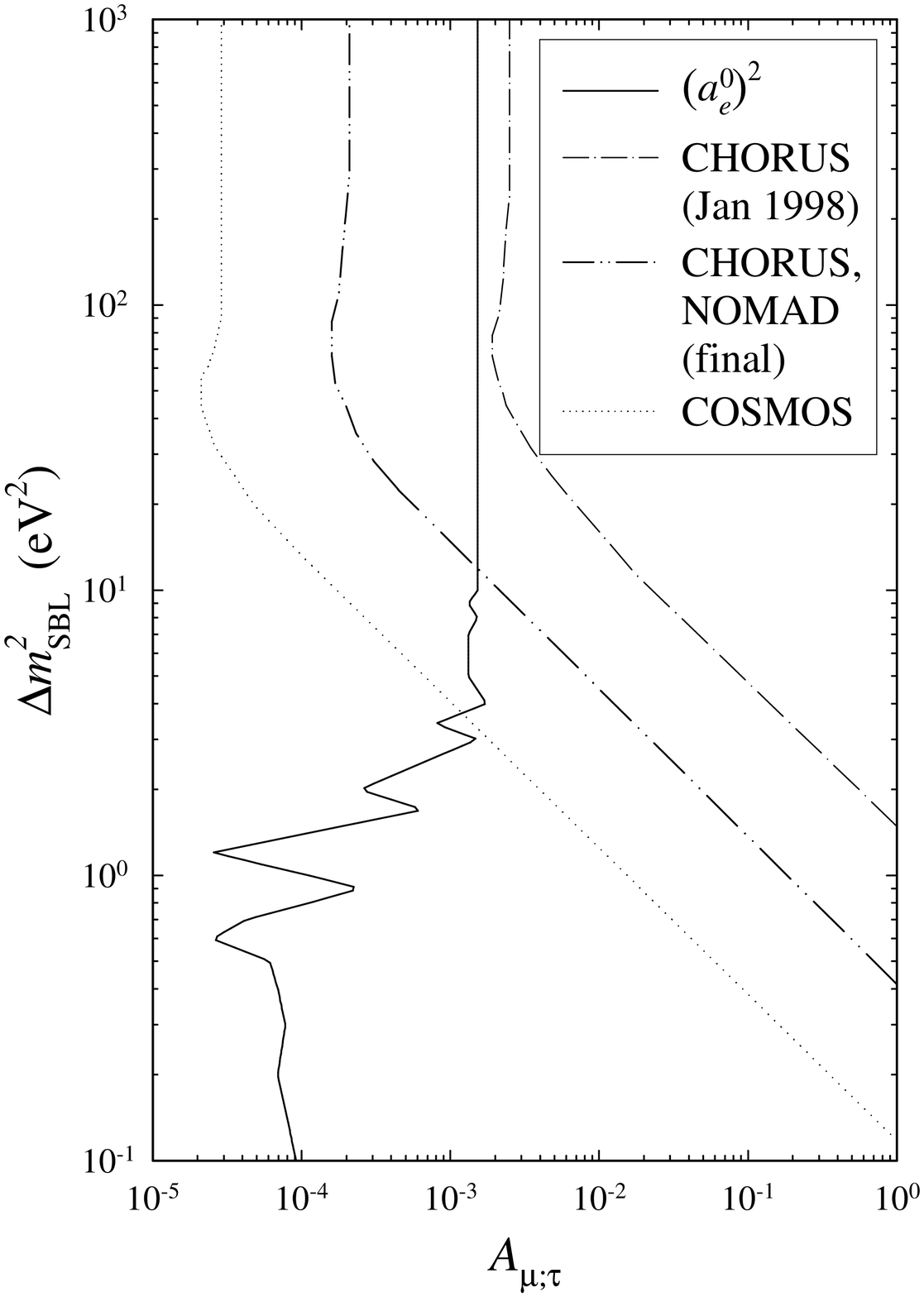,height=0.95\textheight}}
\end{center}
\end{minipage}
\begin{center}
\Large Figure \ref{fig3}
\end{center}

\end{document}